\documentclass[reprint,
superscriptaddress,
nofootinbib,
 amsmath,amssymb,
]{revtex4-2}

\usepackage{natbib,hyperref}

\usepackage{graphicx}
\usepackage{subfigure}
\usepackage{dcolumn}
\usepackage{bm}
\usepackage{hyperref}

\usepackage{xcolor}
\usepackage{subcaption}
\usepackage{autonum}

\renewcommand{\t}[1]{\tilde{#1}}

\begin{document}

\title{Spherically symmetric electrically counterpoised dust either collapses or disperses}
\author{Hugo Visaguirre}
\altaffiliation{hvisaguirre@mendoza-conicet.gob.ar}
\affiliation{Instituto Interdisciplinario de Ciencias B\'asicas, CONICET, Mendoza, Argentina.}
\author{Andr\'es Ace\~na}
\altaffiliation{acena.andres@conicet.gov.ar}
\affiliation{Instituto Interdisciplinario de Ciencias B\'asicas, CONICET, Mendoza, Argentina.}
\affiliation{Facultad de Ciencias Exactas y Naturales, Universidad Nacional de Cuyo, Mendoza, Argentina.}



\begin{abstract}
 We explore the dynamical evolution of spherically symmetric objects made of electrically counterpoised dust in general relativity. It has been claimed that these objects are in neutral equilibrium and, therefore, that black hole mimickers made of electrically counterpoised dust are feasible. Here we show that if a velocity is imparted to the fluid elements, no matter how small, the evolution leads either to a black hole or to the dispersion of the fluid. Furthermore, in the case of collapse, the resulting object is necessarily an extremal Reissner–Nordstr\"om black hole.
\end{abstract}

\maketitle

\section{Introduction}

In this work, we study numerically the dynamical evolution of spherically symmetric configurations of electrically counterpoised dust (ECD) in general relativity (GR). This matter model consists of a charged perfect fluid without pressure, with the charge and mass densities perfectly balanced, which makes static solutions possible. The existence of static spacetimes that satisfy the Einstein-Maxwell system of equations for collections of discrete particles was first shown by Majumdar \cite{majumdar1947class} and Papapetrou \cite{papapetrou1945static}, following the work of Weyl on axisymmetric spacetimes \cite{weyl1917}. If the matter content is restricted to said particles, then to each particle there is an event horizon, which is interpreted as an extremal Reissner-Nordstr\"om (ERN) black hole \cite{hartlehawking1972}. If the main interest is on regular objects, then the exterior solution can be matched with static interiors made of ECD \cite{das1962class}, \cite{varela2003}. The assumptions made in \cite{weyl1917}, \cite{majumdar1947class}, \cite{papapetrou1945static} and \cite{das1962class} have been relaxed in several ways, and the results extended to charged perfect fluids with pressure, for example in \cite{hernandez1967}, \cite{Guilfoyle1999}, \cite{Lemos2017}, or to higher dimensions \cite{lemos2008bonnor}, \cite{Lemos2009}.

If the matter model is ECD, then any static fluid distribution gives rise to a solution of the Einstein-Maxwell system of equations, and this has been used to test physical and geometric possibilities within GR by constructing tailored spacetimes. As examples, we can mention the study of the relationship between charge and mass in the Reissner-Nordstr\"om solution and the construction of a point charge model \cite{bonnor1960mass}, the construction of static objects with unbounded density \cite{bonnor1972static}, to show that unbounded redshifts can be obtained from regular objects \cite{bonnor1975very}, and to discuss the hoop conjecture \cite{bonnor1998model}. Interestingly, the engineered solutions can be as close to the ERN black hole as desired, and this has been analyzed in relation to the bifurcation of solutions \cite{horvat2005regular} and, also, it has been shown that such black hole limit is a general feature of ECD solutions \cite{meinel2011}. This means that a regular ECD object could mimic an ERN black hole as well as desired, and bears importance for the existence of black hole mimickers, as theoretically it is possible to construct them from a relatively normal gas.

Nevertheless, having a fine-tuned matter model together with a delicate limit means that stability considerations are paramount. If the solutions constructed are unstable, then there is no expectation of finding them in nature, as they would always be subjected to some perturbations. Regarding ECD solutions, in general they are considered to be in neutral equilibrium, as there is no force imbalance that would take one solution into another \cite{omote1974criterion,lemos2023stability}. But this does not take into account that a perturbed solution would have an evolution that may not resemble the unperturbed configuration. In the linear regime, for the spherically symmetric case,  it was shown in \cite{Acena:2020yvj} that perturbations to a static ECD solution travel at constant speed. This is a reflection of said indifferent equilibrium, but it was also shown that it is possible to start with a regular solution and end up with a black hole via the perturbation.

Closely related to the question of stability of ECD is the stability of charged fluid spheres with pressure, specially those that are close to extremality in their charge to mass ratio. In \cite{Anninos2001} it was shown that before the spheres reach the ERN limit there is a transition from stable to unstable solutions and, therefore, collapse occurs before the ERN limit is reached. In other words, the collapse of charged spheres can only produce sub-extremal black holes. On the other hand, in \cite{lemos2014} incompressible charged solutions with proportional charge and matter densities are discussed, and they present regular objects on the limit of being black holes. Further considering this issue, relativistic polytropes were studied in \cite{acena2024}, showing that stable and causal black hole mimickers can not be formed from this type of matter. The limit to obtain a black hole mimicker starting with a polytrope is subtle and ends up being ECD. The conclusion is that if it is possible to create black hole mimickers from a charged perfect fluid, it needs to be ECD.

In order to see if the results in \cite{Acena:2020yvj} are just an artifact of the linear approximation, here we consider the coupled Einstein-Maxwell equations and evolve them numerically. Our main conclusion is that the spherically symmetric evolution would lead either to the dispersion of the dust or to its collapse, forming an ERN black hole. This takes the previous conclusions into the non-linear regime and confirms that black hole mimickers made of ECD are not feasible.

The article is organized as follows. In Section \ref{section system of eq} we present the system of equations with ECD as matter model and the criterion for indicating the collapse of matter. The numerical scheme used to evolve suitable initial data is described in Section \ref{section Numerical scheme}. In Sections \ref{Section spherical shell} and \ref{Section Black hole mimicker} we present the numerical evolution performed for two cases of interest. The first one corresponds to a shell of matter collapsing or dispersing, and the second one to the collapse of a black hole mimicker, where we show that a black hole mimicker is unstable to the perturbation. The conclusions are in Section \ref{Section Conclusions}.

\section{The system of equations}\label{section system of eq}

Here we present the system of equations to be considered, namely, the Einstein-Maxwell equations with ECD as matter model. We use geometrized units: $G=c=1$ and $\epsilon_0 = (4\pi)^{-1}$. The equations are
\begin{equation}\label{eq: Einstein-Maxwell}
    G_{\mu\nu} = 8\pi T_{\mu\nu},\quad \nabla_\nu F^{\mu\nu} = 4\pi j^{\mu}, \quad \nabla_{[\mu}F_{\nu\lambda]} = 0,
 \end{equation}
being $G_{\mu\nu}$ the Einstein tensor, $T_{\mu\nu}$ the
energy-momentum tensor, $\nabla_\mu$ the covariant derivative, $F_{\mu\nu}$ the Faraday tensor and $j^\mu$ the current density. $T_{\mu\nu}$ has contributions from the distribution of matter, which is dust, and from the electromagnetic field,
\begin{equation}\label{eq: Energy Tensor}
    T_{\mu\nu} = \rho\, u_\mu u_\nu + \frac{1}{4\pi}\left(F_{\lambda\mu}F^\lambda\,_\nu - \frac{1}{4} F_{\lambda\rho}F^{\lambda\rho}g_{\mu\nu}\right),
\end{equation}
where $\rho$ is the mass density and $u^\mu$ the four-velocity of the fluid. The current density is given in terms of the charge density, $\sigma$,
\begin{equation}
    j^\mu = \sigma u^\mu,
\end{equation}
and the constitutive relation for ECD is
\begin{equation}
    \sigma = \rho.
\end{equation}

We restrict our analysis to spherical symmetry. Then, in adapted coordinates, the metric can be written as
\begin{align}\label{métrica}
    ds^2 = & - \Phi(r,t)^2 dt^2 + \Lambda(r,t)^2 dr^2 \\
    & + r^2 \Psi(r,t)^2 (d\theta^2+\sin^2\theta\, d\phi^2).
\end{align}
We use the coordinate freedom to make the $r$ coordinate identify each fluid element, that is,
\begin{equation}
    u^\mu = \Phi^{-1}\,\partial_t^\mu.
\end{equation}
Due to spherical symmetry, the only nonzero components of the Faraday tensor are $F_{tr} = -F_{rt}$ and $F_{\theta\phi}=-F_{\phi\theta}$. As the solutions need to be regular at the origin, $F_{\theta\phi}$ is also zero, and then
\begin{equation}
    F_{\mu\nu} = 2E(r,t)\,dt_{[\mu} dr_{\nu]},
\end{equation}
where $E(r,t)$ is the radial component of the electric field.

Charge is conserved, and we are using coordinates comoving with the fluid, this implies that the total charge inside $r$ is independent of $t$,
\begin{equation}
    Q(r) = 4\pi \int_0^r \sigma(r,t) \Lambda(r,t)\Psi(r,t)^2 r^2 dr.
\end{equation}
Therefore, it is convenient to use $Q$ as a variable instead of $\sigma$ and $\rho$,
\begin{equation}\label{eq rho}
    \rho = \sigma = \frac{\partial_r Q}{4 \pi r^2 \Lambda \Psi^2}.
\end{equation}

We write the system of equations as a first order PDE system, defining $\Phi_t$, $\Psi_r$, $\Psi_t$ and $\Lambda_t$ through
\begin{align}
    \partial_t\Phi & = \Phi_t, \label{dot Phi} \\
    \partial_r\Psi & = \Psi_r, \label{eq Psi'} \\
    \partial_t\Psi & = \Psi_t, \label{dot Psi} \\
    \partial_t\Lambda & = \Lambda_t. \label{eq dotLambda}
\end{align}
Then, the Einstein-Maxwell system of equations for ECD in spherical symmetry is
\begin{align}
    \partial_r\Phi = & \frac{Q\,\Lambda \, \Phi}{r^2\Psi^2}, \label{eq Phi'}\\ 
    \partial_r\Phi_t = & \frac{Q}{r^2\Psi^3}\left(\Phi \Psi \Lambda_t + \Lambda \Psi \Phi_t - 2 \, \Lambda \Phi \Psi_t\right), \label{eq dotPhi'} \\
    \partial_r\Psi_r = & \frac{\Psi}{r} \left(\frac{ \partial_r\Lambda}{\Lambda}-\frac{3 \Psi_r}{\Psi}\right) -\frac{\Psi}{2 r^2}\left(1- \frac{\Lambda^2}{\Psi^2}\right) \label{eq Psi''}\\
    & + \frac{\partial_r\Lambda \Psi_r}{\Lambda} - \frac{\Psi_r^{2}}{2 \Psi} + \frac{\Lambda \Lambda_t \Psi_t}{\Phi^{2}} + \frac{\Lambda^{2} \Psi_t^{2}}{2 \Phi^{2} \Psi} \\
    & - \frac{\partial_rQ \Lambda}{r^2 \Psi} - \frac{Q^2 \Lambda^{2}}{2 r^4 \Psi^{3}}, \\
    \partial_r\Psi_t = & \frac{\Psi}{r } \left(\frac{ \Lambda_t}{\Lambda} -\frac{\Psi_t}{\Psi}\right) + \frac{\Lambda_t \Psi_r}{\Lambda} + \frac{Q\Lambda \Psi_t}{r^2 \Psi^{2}}, \label{eq dotPsi'} \\
    \partial_t\Psi_t = & \frac{\Phi^{2} \Psi_r}{r \Lambda^{2}} + \frac{\Phi^2\Psi}{2 r^2 \Lambda^2}\left(1-\frac{\Lambda^2}{\Psi^2}\right) + \frac{\Phi^{2} \Psi_r^{2}}{2 \Lambda^{2} \Psi} \label{eq ddotPsi}\\
    & + \frac{\Phi_t \Psi_t}{\Phi} - \frac{\Psi_t^{2}}{2 \Psi} +  \frac{Q \Phi^{2} \Psi_r}{r^2\,\Lambda \Psi^{2}} \\
    & + \frac{Q \Phi^{2} }{r^3\Lambda \Psi} + \frac{Q^2 \Phi^{2}}{2 r^4 \Psi^{3}}, \\
    \partial_t\Lambda_t = & -\frac{2 \Phi^{2} \Psi_r}{r \Lambda \Psi}-\frac{\Phi^2}{\Lambda r^2}\left(1-\frac{\Lambda^2}{\Psi^2}\right) - \frac{\Phi^{2} \Psi_r^{2}}{\Lambda \Psi^{2}} \label{eq ddotLambda}\\
    & - \frac{\Lambda_t \Phi_t}{\Phi} + \frac{\Lambda \Psi_t^{2}}{\Psi^{2}} - \frac{2Q \Phi^{2} \Psi_r}{ r^2 \Psi^{3}} \\
    & - \frac{2Q \Phi^{2}}{ r^3\Psi^{2}} - \frac{Q^2\Lambda \Phi^{2}}{r^4\Psi^{4}}.
\end{align}

Due to the use of spherical coordinates, the system of equations is formally singular at $r=0$, so we impose regularity conditions. Firstly, the different variables need to be well-behaved at the origin, which means that the Taylor expansion of the metric functions and the function $Q$ are
\begin{align}
&\Phi = \Phi_0 + \Phi_2 r^2 + \mathcal{O}(r^3),\\
&\Lambda = \Lambda_0 + \Lambda_2 r^2 + \mathcal{O}(r^3),\\
&\Psi = \Psi_0 + \Psi_2 r^2 + \mathcal{O}(r^3),\\
&Q = r^3\,Q_0 + \mathcal{O}(r^4),
\end{align}
with $\Phi_0$, $\Phi_2$, $\Lambda_0$, $\Lambda_2$, $\Psi_0$ and $\Psi_2$ functions that may depend on $t$ but not on $r$ and $Q_0$ a constant. On the other hand, the spacetime must be locally flat at $r=0$, which implies that
\begin{equation}
    \Psi_0(t) = \Lambda_0(t).
\end{equation}

The system of equations has a scaling invariance, that is, it has the same form if we make the replacements
\begin{gather}
    Q \to a Q,\quad t \to b t, \quad r \to c r,\\
    \Lambda \to \frac{a}{c} \Lambda,\quad \Psi \to \frac{a}{c} \Psi, \quad \Phi \to \frac{a}{b}\Phi,
\end{gather}
with $a$, $b$ and $c$ positive constants, and the corresponding rescaling for the derivatives. Also, there is the time-reversal symmetry,
\begin{equation}
    t \to -t.
\end{equation}
We are interested in the interior solution, that is, we consider that $\rho\neq 0$ for $0\leq r \leq R$ and that $\rho =0$ for $r>R$. It is convenient to use dimensionless variables, making the replacements
\begin{equation}
    r \to \frac{r}{R},\quad t \to \frac{t}{R}, \quad Q \to \frac{Q}{R}.
\end{equation}
The system of equations is the same as before, but now the domain for the interior solution is $0\leq r\leq 1$. The exterior solution is directly the ERN spacetime, which is glued to the interior solution using the Israel-Darmois juncture conditions.

From the analysis in \cite{Acena:2020yvj} we expect that, depending on the initial data, the solution may disperse or collapse forming a black hole. If the matter disperses, the spacetime simply approaches the Minkowski spacetime while the matter density goes to zero. On the other hand, in order to see locally in time if the matter is collapsing into a black hole, we check for the appearance of apparent horizons. If we take the hypersurface $t=constant$, and in it the sphere $r=constant$, then the expansion of the corresponding outgoing null geodesics is given by
\begin{equation}
    H = -2\frac{\left(r\Phi\Psi_r + r \Lambda\Psi_t + \Phi\Psi\right)}{r\Lambda\Phi\Psi}.
\end{equation}
We have an apparent horizon whenever $H = 0$.

\section{Numerical scheme}\label{section Numerical scheme}

We have an overdetermined system of equations for the unknowns $\Phi$, $\Phi_t$, $\Psi$, $\Psi_r$, $\Psi_t$, $\Lambda$ and $\Lambda_t$. We use a fully constrained integration scheme where in each time step we first solve the constraint equations \eqref{eq Phi'}, \eqref{eq dotPhi'}, \eqref{eq Psi'}, \eqref{eq Psi''} and \eqref{eq dotPsi'} to obtain $\Phi$, $\Phi_t$, $\Psi$, $\Psi_r$ and $\Psi_t$. Then we evolve one step in time using equations \eqref{eq dotLambda} and \eqref{eq ddotLambda} for $\Lambda$ and $\Lambda_t$. Equations \eqref{dot Phi}, \eqref{dot Psi} and \eqref{eq ddotPsi} are left as control equations. With this separation of the system of equations, we have one free function, $Q(r)$, and two functions of $r$ as initial data,
\begin{equation}
    \Lambda^0(r) = \Lambda(r,0),\quad \Lambda^0_t(r) = \Lambda_t(r,0).
\end{equation}
For the integration of the constraints we need the boundary value at $r=0$ of the corresponding functions, that is, $\Phi(0,t)$, $\Phi_t(0,t)$, $\Psi(0,t)$, $\Psi_r(0,t)$ and $\Psi_t(0,t)$. Due to the condition of local flatness we have
\begin{equation}
    \Psi(0,t) = \Lambda(0,t),\,\, \Psi_r(0,t)= 0, \,\, \Psi_t(0,t) = \Lambda_t(0,t).
\end{equation}
On the other hand, $\Phi(0,t)$ can be given freely, determining $\Phi_t(0,t)$ through \eqref{dot Phi}, and we choose
\begin{equation}
    \Phi(0,t) = 1,\quad \Phi_t(0,t) = 0,
\end{equation}
so that the time coordinate $t$ at the origin is the proper time of the corresponding fluid element.

A particular feature of the system of equations is that if we choose $\Lambda_t^0(r)=0$, then the solution is automatically static. This is due to the matter model being ECD, which means that there is no fundamental equilibrium configuration that can be singled out. This also means that we can always compare the dynamical spacetime to a static one. The static spacetime is obtained in the following way \cite{Acena:2020yvj}: we start by choosing $Q(r)$, then
\begin{equation}
    \Psi_{s} = 1, \quad \Lambda_{s} = \frac{r}{r-Q},
\end{equation}
and $\Phi_{s}$ is obtained solving \eqref{eq Phi'} with the boundary condition $\Phi_{s}(0)=1$, where the subscript $s$ indicates that we are referring to the static background solution. Furthermore, we use the static solution to avoid dealing with the formal singularity at $r=0$.  Instead of integrating the system of equations in the range $0\leq r \leq 1$, we choose a value of $r_0$, close to $0$ and integrate in the range $r_0 \leq r \leq 1$. We impose the boundary conditions for the integration of the constraints given by the static solution:
\begin{align}
    & \Psi(r_0,t) = 1,\quad \Psi_r(r_0,t) = 0, \quad \Psi_t(r_0,t) = 0 , \\
    & \Phi(r_0,t) = \Phi_{s}(r_0), \quad \Phi_t(r_0,t) = 0.
\end{align}
Also, we choose the initial data for $\Lambda$ as
\begin{equation}
    \Lambda^0(r) = \Lambda_s(r).
\end{equation}
Then, the solution for $0\leq r \leq r_0$ is the static background solution, and we do not need to solve the system of equations there. For $r_0\leq r\leq 1$, we give $\Lambda_t^0(r)$ as a Gaussian profile:
\begin{equation}\label{Lambdat0 inicial}
    \Lambda_{t}^0 =  A e^{-\frac{(r-r_c)^2}{\sigma^2}}.
\end{equation}
Here $A$ is the amplitude, which can have either sign, $r_c$ is the center of the Gaussian and $\sigma$ its width. We interpret $\Lambda_t^0$ as imparting a velocity to the fluid elements, which corresponds to the coordinate velocity
\begin{equation}\label{velocidad}
    v(r,t) = \int_0^r \Lambda_t(\t{r},t)\,d\t{r}.
\end{equation}
We see that if we use the scaling invariance $t\to bt$, then
\begin{equation}
    \Lambda_t \to \frac{1}{b}\Lambda_t,\quad v\to \frac{1}{b}v.
\end{equation}
This means that changing the absolute value of $A$ only impacts on the speed at which the evolution takes place, but it does not change the evolution itself. In particular, the final fate of the evolution, be it dispersion or collapse, is the same regardless the value of $|A|$. Furthermore, due to the time reversal symmetry, the sign of $A$ determines if the solution is collapsing or dispersing, and one solution can be turned into the other by changing said sign.

We built a numerical code that integrates the constraint equations \eqref{eq Psi'}, \eqref{eq Phi'}, \eqref{eq dotPhi'}, \eqref{eq Psi''} and \eqref{eq dotPsi'} in space at each time step and then evolves the equations for $\Lambda$ and $\Lambda_t$, \eqref{eq dotLambda} and \eqref{eq ddotLambda}, one step in time. The entire numerical algorithm is built in Julia \cite{bezanson2017julia}, the spatial integrations of the constraints are performed with the Differential Equations (DE) package \cite{rackauckas2017differentialequations}, using a fourth-order Runge Kutta algorithm (RK4). The obtained data is used as input for the time integrations which are made with an explicit RK4. At each time step the results for $\Lambda$ and $\Lambda_{t}$ are interpolated to be used in the DE integrator. We use a grid spacing  $\Delta r = 0.001$  and the time step $\Delta t$ is chosen appropriately for each case, depending on the total time of integration and the value of $|A|$ in order for the pulse not to be lost between time steps. As a check of accuracy, we have evolved known static solutions and the numerical errors are correctly bounded.

In the next sections, we present the numerical evolution performed for the two cases that we are interested in. The first one corresponds to a shell of matter collapsing or dispersing, and the second one to the collapse of a black hole mimicker.

\section{Spherical shell}\label{Section spherical shell}

\begin{figure}
  \begin{center}
    \subfigure[Metric function $\Phi$.]{
        \includegraphics[width=0.45\textwidth]{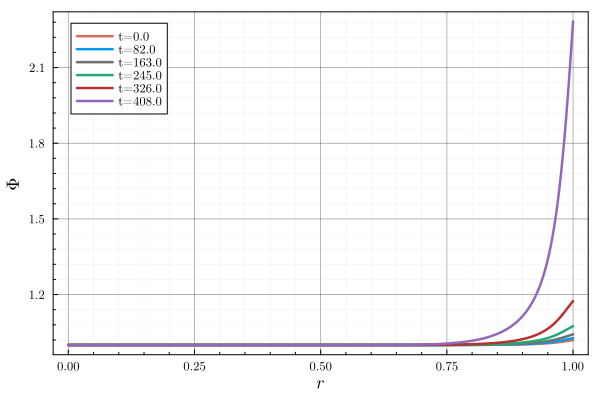}
        \label{fig evol Phi colapso}}    
    \subfigure[Metric function $\Lambda$.]{
         \includegraphics[width=0.45\textwidth]{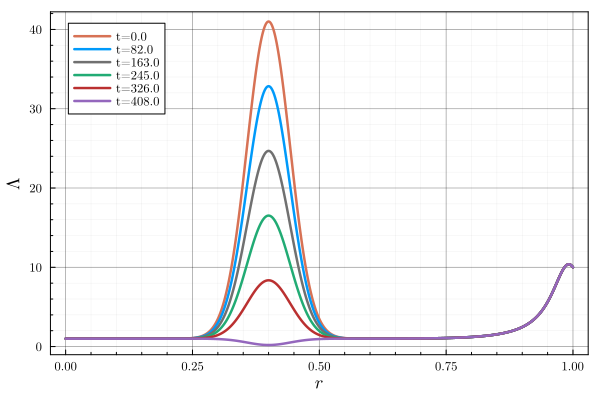}
        \label{fig evol Lambda colapso}}
        \subfigure[Metric function $\Psi$.]{
         \includegraphics[width=0.45\textwidth]{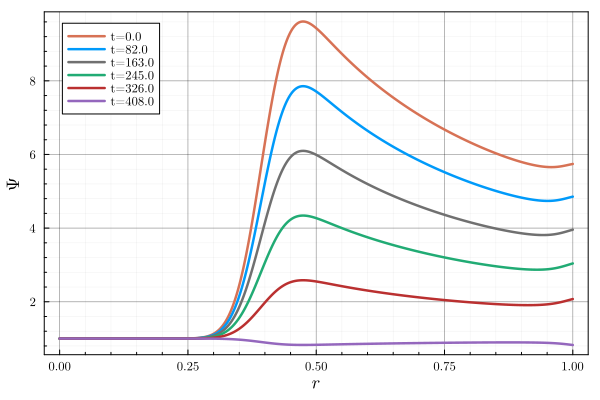}
        \label{fig evol Psi colapso}}
    \caption{Evolution of the metric functions for the case of the spherical shell.}
    \label{fig evol metric function colapso}
  \end{center}
\end{figure}

\begin{figure}
    \centering
    \includegraphics[width=\linewidth]{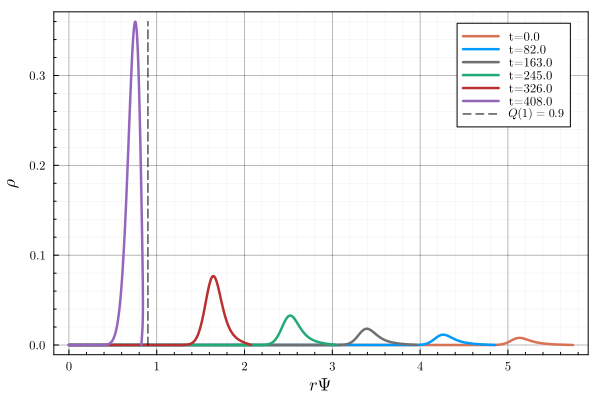}
    \caption{Evolution of the density $\rho$ of the spherical shell in areal radius. It can be seen how the object is becoming more compact and eventually crosses the limit $r \Psi= 0.9$, represented by a black dash line.}
    \label{fig:rho vs xarea}
\end{figure}

\begin{figure}
    \centering
    \includegraphics[width=\linewidth]{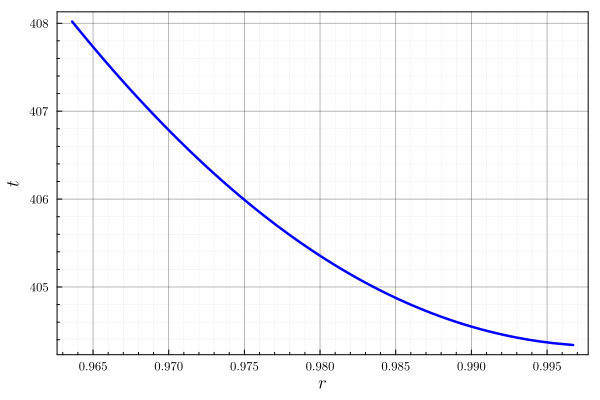}
    \caption{Coordinate location of the apparent horizon. As the ECD shell is becoming more compact it is forming an apparent horizon indicating that the end result is a black hole.}
    \label{fig:Horiz_Apar_r_vs_t}
\end{figure}

In this section we investigate the prototypical spherical collapse. That is, we start with a matter distribution that resembles a spherical shell and that is far from being a black hole. For the evolution we give a small and practically equal inward velocity to all the fluid elements. To address this case, we take advantage of the time-reversal symmetry of the system of equations. We can evolve forward and backward the system, as long as a collapse does not occur. Therefore, we construct static initial data representing a spherical shell and evolve it outward until we get initial data representing an object far from forming a black hole. Here, since a sufficiently large time is integrated, we use $\Delta t =0.02$. We can construct a spherical shell using a function $Q$ of the form
\begin{equation}
    Q = \frac{9}{2560} r^8 \left(6-5r\right)^4 \left(1-3r\right)^8, 
\end{equation}
where $Q(1) = 0.9$ is the value of the total charge. The initial data $\Lambda_t^0$ is given by \eqref{Lambdat0 inicial} with parameters $|A|= 0.1$, $r_c = 0.4$ and $\sigma = 0.06$. Next, $\Lambda^0$ is given by the static solution plus a Gaussian. This is the form of $\Lambda$ when we evolve with $\Lambda_t^0$ with $A$ positive until $t=400$  
\begin{equation}
    \Lambda^0 = \frac{r}{r-Q} + 40\,e^{-\frac{(r-r_c)^2}{\sigma^2}},
\end{equation}
 again $r_c = 0.4$ and $\sigma = 0.06$. With this initial condition, the static background in the region $0\leq r \leq r_0$ is Minkowski, i.e. the boundary imposed in $r_0$ is Minkowski. Using $A$ negative, i.e. giving an inward velocity, the evolution goes until the function $\Lambda$ vanish at $r=r_c$ and the integrator collapse. Figure \ref{fig evol metric function colapso} shows the evolution of the metric functions $\Phi$, $\Lambda$ and $\Psi$, where it can be seen how the function $\Lambda$ goes to zero at $r=r_c$. As we evolve with an inward velocity, we see that the velocity of the shell is constant, and the object is becoming more compact. Figure \ref{fig:rho vs xarea} shows the evolution of the density $\rho$ versus the areal radius, showing how the object is becoming more compact crossing the critical value $r\Psi = 0.9$. At the same time, an apparent horizon is forming. The apparent horizon was computed using a zero-ﬁnding algorithm, which gives us the coordinates where $H=0$. Figure \ref{fig:Horiz_Apar_r_vs_t} shows the apparent horizon in coordinate radius. Due to the formation of the apparent horizon, we have that once the pulse starts moving inward, the electric repulsion is not enough to stop it, and it ends in an inevitable collapse, forming a black hole. Given the scaling invariance, we have that if the value of $A$ is negative, the final fate of the matter shell is to collapse regardless the magnitude of $|A|$. So we can say that if we give an initial inward velocity to the spherical shell, we see numerically that apparent horizons are formed, therefore, the shell collapses forming a black hole.

\section{Black hole mimicker}\label{Section Black hole mimicker}

\begin{figure}
  \begin{center}
    \subfigure[Metric function $\Phi$.]{
        \includegraphics[width=0.45\textwidth]{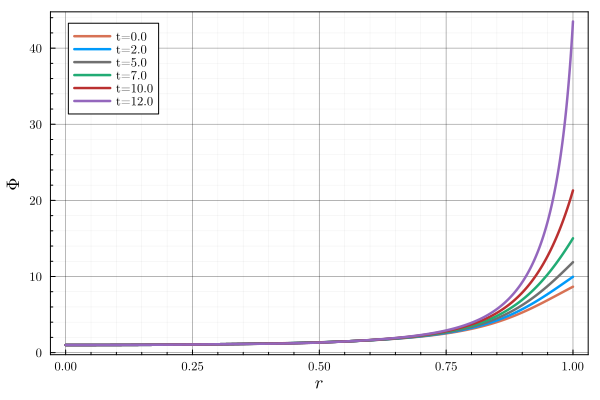}
        \label{fig evol Phi QBH}}
    \subfigure[Metric function $\Lambda$.]{
        \includegraphics[width=0.45\textwidth]{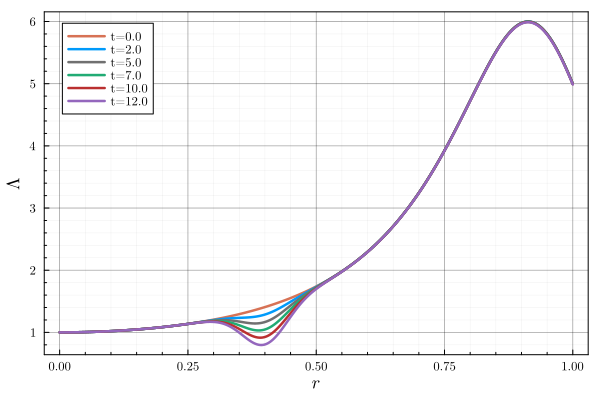}
        \label{fig evol Lambda QBH}}
    \subfigure[Metric function $\Psi$.]{
        \includegraphics[width=0.45\textwidth]{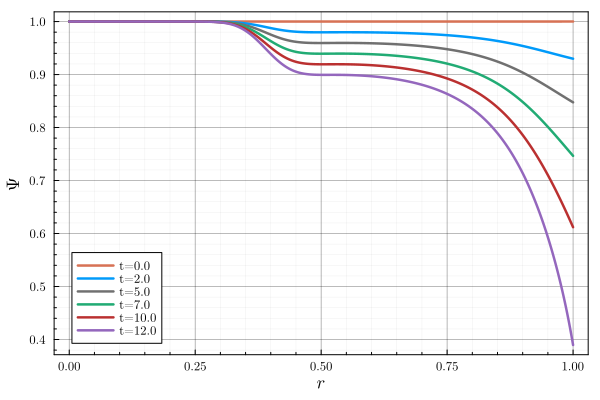}
        \label{fig evol Psi QBH}}
    \caption{Evolution of the metric functions for the case of the black hole mimicker.}
    \label{fig evol metric function QBH}
  \end{center}
\end{figure}

\begin{figure}
    \centering
    \includegraphics[width=\linewidth]{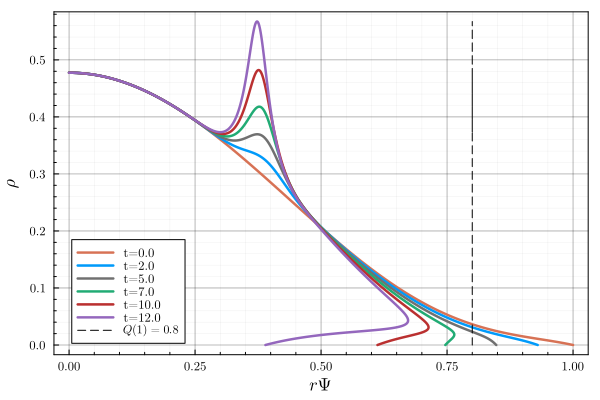}
    \caption{Evolution of the density $\rho$ of black hole mimicker in areal radius. It can be seen how the edge is going inward making the mimicker more compact and at some time it crosses the limit $r \Psi= 0.8$, represented by a black dash line.}
    \label{fig:rho vs xarea_QBH}
\end{figure}

\begin{figure}
    \centering
    \includegraphics[width=\linewidth]{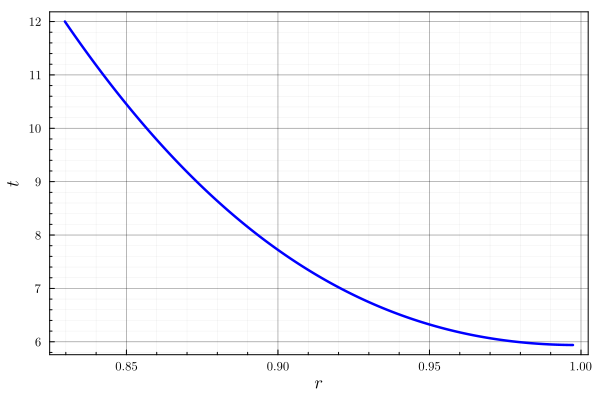}
    \caption{Coordinate location of the apparent horizon. As the black hole mimicker is becoming more compact it is forming an apparent horizon indicating that the end result is a black hole.}
    \label{fig:Horiz_Apar_r_vs_tpro_QBH}
\end{figure}

The second case that we consider is that of a black hole mimicker being perturbed. For this, we use the results of \cite{Acena:2020yvj}, where they present such an object. The initial conditions are given by
\begin{equation}
 Q =   \frac{2}{5}r^3\left(5-3r^2\right),
\end{equation}
and 
\begin{equation}
    \Lambda^0 = \frac{1}{1-\frac{2}{5} r^2(5-3r^2)},
\end{equation}
with $Q(1) = 0.8$ the total charge. Again $\Lambda_t^0$ is given by \eqref{Lambdat0 inicial}, but now, with parameters $A= -0.05$, $r_c = 0.4$ and $\sigma = 0.06$. For this simulation, $\Delta t =\frac{\Delta r}{2}=0.0005$. The simulation goes until we reach a final time $t=12$, not being able to evolve beyond this time due to the collapse of the integrator. Figure \ref{fig evol metric function QBH} shows how the inward velocity impact on the metric functions. In this case we see that $\Lambda$ does not go to zero, $\Phi$ grows  and $\Psi$ is decreasing, in consequence the proper area $r\Psi$ is also decreasing. In this case, when integrating the velocity, we see that it is not affecting the entire matter distribution. As expected from the construction of the initial data there is a region near the edge moving to the left and another near the origin that remains static. Still, as the edge is moving to the left, the object is becoming more compact. Figure \ref{fig:rho vs xarea_QBH} shows density vs areal radius, and we see how the edge is moving to the left and crosses the limit $r\Psi=0.8$, obtaining a more compact object. As we did before, we find the points that indicate the formation of an apparent horizon using the zero-ﬁnding algorithm. Figure \ref{fig:Horiz_Apar_r_vs_tpro_QBH} present the location of the apparent horizon in coordinate radius vs time. This tells us that, if we give an inward velocity, as small as we want, the object will become more compact and eventually collapse forming a black hole. We conclude that the black hole mimicker is not stable in the face of a small perturbation.

\section{Conclusions}\label{Section Conclusions}

In this work, we present the dynamical equations for ECD in spherical symmetry and set up a fully constrained numerical integrator to evolve two initial data representing cases of interest. The first, a spherical shell to recreate a typical case of gravitational collapse, where the shell can be given an initial velocity either inward or outward. We found that the object can be pushed away with an outward velocity and then if the sign of the velocity is changed causing the shell to move inward, no matter the magnitude of the velocity, the final result of the evolution is the formation of a black hole. In the second case, a black hole mimicker, we show that if we perturb it with an inward velocity, it becomes more compact and apparent horizons appear indicating the collapse and formation of a black hole. This tells us that a black hole mimicker made of ECD is unstable to perturbations. Therefore, the main conclusion is that static ECD spacetimes are not stable in the sense that a dynamical perturbation will take the spacetime away from the first static configuration. The possible outcomes are dispersion or collapse into a black hole, or a combination of both, but no return to the original configuration or settling down into another regular static configuration. This, together with the fine-tuning necessary at the level of the matter model, makes it hard to consider that ECD objects are feasible candidates as black hole mimickers.

%
%
%
%
%


\bibliography{Evol_ECD}
\bibliographystyle{abbrvnat}

\end{document}